\documentclass[twocolumn, prb, showpacs, amsmath, amssymb, superscriptaddress]{revtex4}
\usepackage{graphicx}% Include figure files
\usepackage{amsmath}

\begin{document}

\title{Spin effects in single-electron transport through carbon nanotube quantum dots}% Force line breaks with \\

\author{Satoshi Moriyama}
\altaffiliation{Present address: International Center for Young Scientists (ICYS), National Institute for Materials Science (NIMS), 1-1, Namiki, Tsukuba, Ibaraki 305-0044, Japan
}%
\email{MORIYAMA.Satoshi@nims.go.jp}
\affiliation{
Advanced Device Laboratory, The Institute of Physical and Chemical Research (RIKEN), 2-1, Hirosawa, Wako, Saitama 351-0198, Japan
}%
\author{Tomoko Fuse}
\affiliation{
Advanced Device Laboratory, The Institute of Physical and Chemical Research (RIKEN), 2-1, Hirosawa, Wako, Saitama 351-0198, Japan
}%
\author{Tomohiro Yamaguchi}
\affiliation{
Advanced Device Laboratory, The Institute of Physical and Chemical Research (RIKEN), 2-1, Hirosawa, Wako, Saitama 351-0198, Japan
}%
\author{Koji Ishibashi}
\affiliation{
Advanced Device Laboratory, The Institute of Physical and Chemical Research (RIKEN), 2-1, Hirosawa, Wako, Saitama 351-0198, Japan
}%
\affiliation{
CREST, Japan Science and Technology (JST), Kawaguchi, Saitama 332-0012, Japan
}%

\date{\today}% It is always \today, today,
             %  but any date may be explicitly specified

\begin{abstract}
We investigate the total spin in an individual single-wall carbon nanotube quantum dot with various numbers of electrons in a shell by using the ratio of the saturation currents of the first steps of Coulomb staircases for positive and negative biases. The current ratio reflects the total-spin transition that is increased or decreased when the dot is connected to strongly asymmetric tunnel barriers. Our results indicate that total spin states with and without magnetic fields can be traced by this method.
\end{abstract}

% PACS, the Physics and Astronomy
                             % Classification Scheme.
%\keywords{Suggested keywords}%Use showkeys class option if keyword
                              %display desired
\maketitle

%\pagebreak

The electron-spin configurations in quantum dots are important not only for understanding electron-electron interactions but also for the realization of spin-based quantum computing devices in solid state systems~\cite{spin01}. 
Recently, spin states in carbon nanotube quantum dots have been revealed in various systems with the Kondo effect~\cite{Kondo01, Kondo02, Kondo03} and 
the simple shell structures~\cite{shell01, shell06, shell02, shell03, shell04}.
In single-wall carbon nanotube (SWCNT) quantum dots, two- or four-electron shell structures are predicted, depending on the relation between the zero-dimensional (0D) level spacing ($\Delta$) and an energy mismatch ($\delta$) in the two subbands in the SWCNT~\cite{text01}.
The two-electron shell structure ($\delta \sim \Delta / 2$) originates from the twofold spin degeneracy of single-particle levels, and the four-electron shell structure ($\delta \ll \Delta$) originates from the predicted twofold subband degeneracy in addition to the spin degeneracy~\cite{shell05}. Because the 0D level spacing and charging energy ($E_{C} = e^{2} / C_{\Sigma}$ :$C_{\Sigma}$ is the self-capacitance of a dot) are significantly larger than the interaction energies of the on-site Coulomb energy ($\delta U$) and the exchange interaction energy ($J$), it is possible to maintain the shell structures, regardless of the number of electrons in the dot~\cite{Oreg}.
We have clearly demonstrated the two- and four-electron shell structures in the closed SWCNT 
quantum dots and investigated the total-spin transition by considering the magnetic field evolution of the single-electron transport and by performing excitation spectroscopy measurements~\cite{shell02, APL01}. 

In this study, we carefully investigate the total spin in an individual SWCNT quantum dot by using the ratio of the saturation currents for positive and negative biases in single-electron transport through the dot. 
When a quantum dot is connected to strongly asymmetric tunnel barriers, single-electron tunneling reflects the total-spin transition that is increased or decreased~\cite{Cobden, Akera, Hanson}.
We find that total-spin transitions in magnetic fields, such as the higher-spin states ($S = 1$), can be traced by this method and the experimental results are consistent with the previous results~\cite{shell02}.

The single-electron transport measurements are carried out by applying source-drain bias voltage 
($V_{\text{sd}}$) between left and right barriers. We define the positive polarity of $V_{\text{sd}}$ when electrons flow from the left reservoir to the right reservoir. Then, we define the saturation currents of the first steps of the Coulomb staircases for positive and negative $V_{\text{sd}}$'s as $I_{+}$ and $I_{-}$, respectively. 
When the tunneling rates of the left and right barriers of the dot (represented by $\varGamma_{\text{L}}$ and $\varGamma_{\text{R}}$, respectively) are significantly different, the current is only limited by the thicker barrier with the smaller tunneling rate. For instance, if the left barrier is thicker ($\varGamma_{\text{L}} \ll \varGamma_{\text{R}}$), the current of the single-electron tunneling is approximately determined by $\varGamma_{\text{L}}$~\cite{JPFujisawa}.
In this case, the ratio of saturation current reflects the total spin states in quantum dots, with the relationship $I_{+} / I_{-} = (2 S_{N +1} + 1) / (2 S_{N} + 1)$~\cite{Akera, Hanson}, where $S_{N}$ is the total-spin state with $N$-electrons; $S_{N + 1}$, the total-spin state with $(N + 1)$-electrons. Both $S_{N}$ and $S_{N + 1}$ are for the ground state. It should be noted  that single-particle levels in a quantum dot are degenerate when this argument is applied.
For instance, for the ground state transition from $S_{N} = 0$ to $S_{N + 1} = 1/2$, 
the tunneling rate for positive $V_{\text{sd}}$ is twice that 
for negative  $V_{\text{sd}}$. For positive  $V_{\text{sd}}$, either a spin-up or spin-down electron can enter the dot and exit rapidly due to the larger tunneling rate of the right thinner barrier~\cite{explain02}. On the other hand, for negative $V_{\text{sd}}$, only one electron (spin-up or spin-down) can rapidly enter the dot and exit through the left thicker barrier. 
Therefore, assuming that the tunneling rates are independent of spin states, the current ratio is $I_{+}/I_{-} = 2$. 
For the ground state transition from $S_{N} = 1/2$ to $S_{N + 1} = 0$, 
only one electron (spin-up or spin-down) can enter the dot and create a spin pair in a spin-degenerate single-particle 
level ($S = 0$) and one of the two electrons exits the dot. Since the current is limited by the incoming tunneling rate,  
the current ratio becomes $I_{+} / I_{-} = 1/2$.
The experimental results for this method have been demonstrated in carbon nanotube quantum dots at zero magnetic field~\cite{Cobden}, and in lateral quantum dots made of GaAs/AlGaAs two-dimensional electron gas system~\cite{Hayashi}.

SWCNT quantum dots were fabricated by depositing metallic (Ti) source and drain contacts on top of an individual SWCNT using 50 keV electron beam lithography. 
In our devices, the entire nanotube between the contacts behaved as a single quantum dot~\cite{process01, process02}.
The distance between the contacts was designed to be 300 nm, and a heavily p-doped Si substrate was used for the application of the gate voltage $V_{\text{g}}$. All the electrical transport measurements were carried out in a dilution refrigerator with a superconducting magnet. In our experiments, magnetic fields were applied perpendicular to the tube axis.

\begin{figure}
\includegraphics[width=8cm]{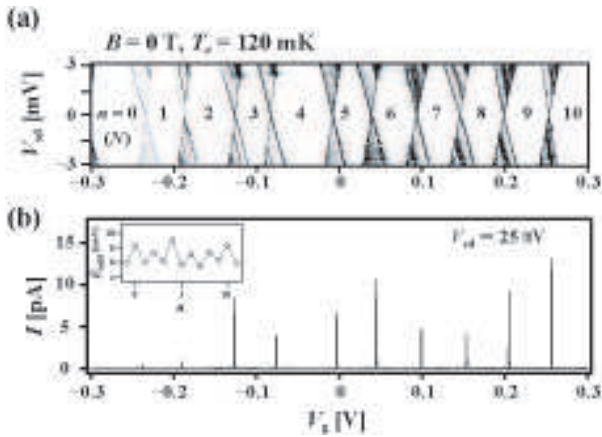}% 
\caption{\label{fig1} (a) Gray scale plot of the differential conductance $dI/dV_{\text{sd}}$ as a function of $V_{\text{sd}}$ and $V_{\text{g}}$ at $B = 0$ T. The number $n$ indicates the number of extra electrons estimated from the Coulomb diamond around $V_{\text{g}} \sim -0.27$ V. $N$ is the total number of electrons in the dot and is suggested to be an even number (see text). (b) Coulomb oscillations in the gate-voltage range that corresponds to (a). Inset: Addition energy ($E_{\text{add}}$) as a function of $n$. The alternate change in the addition energy is observed, indicating 
the even-odd effect~\cite{APL01, Ralph}.}
\end{figure}

Figure 1(a) shows a gray scale plot of the differential conductance $d I / d V_{\text{sd}}$ as a function of $V_{\text{sd}}$ and $V_{\text{g}}$ at $B = 0$ T.
The white-colored regions are Coulomb blockaded regions, and the lines outside the diamonds are due to the effect of zero-dimensional confined levels. 
Figure 1(b) shows the Coulomb oscillations in the region of Fig. 1(a) at a small voltage of $V_{\text{sd}} = 25$ $\mu$V.  
The electron temperature, $T_{e} = 120$ mK ($\sim 10$ $\mu$eV), is obtained by the theoretical fitting of an individual Coulomb peak (not shown here)~\cite{Beenakker, pulse}.
The electrostatic potential of the dot $\phi_{\text{dot}}$ is linearly related to $V_{\text{g}}$ with the standard single-particle model~\cite{text02}. We also deduce that the conversion factor is $\alpha \equiv d \phi_{\text{dot}} / d V_{\text{g}} = 0.15$ meV$/$mV from the fitting of the Coulomb peaks. The distance between adjacent peaks ($\Delta V_{\text{g}}$), in fact, oscillates as a function of the gate voltage with a period of two electrons. 
It corresponds to the addition energy $E_{\text{add}} = \alpha \Delta V_{\text{g}}$. The inset in Fig. 1(b) shows $E_{\text{add}}$ as a function of the number of extra electrons in the dot. The oscillation is clearly observed, indicating the even-odd effect of the number of electrons in the dot~\cite{Ralph}. These results are similar to those of the typical ``closed'' quantum-dot behavior of an individual SWCNT, which has been reported previously~\cite{APL01}.
For larger diamonds, $E_{\text{add}}$ is estimated to be $8-10$ meV when the number of electrons in the dot is even, 
and for the smaller diamonds, it is estimated to be $\sim$ 6 meV when the number of electrons is odd. 
$\delta$ is distributed in the range of $2-3$ meV for different Coulomb diamonds, 
which is estimated from the excitation lines outside the Coulomb diamonds in Fig. 1(a). 
 
\begin{figure}
\includegraphics[width=8cm]{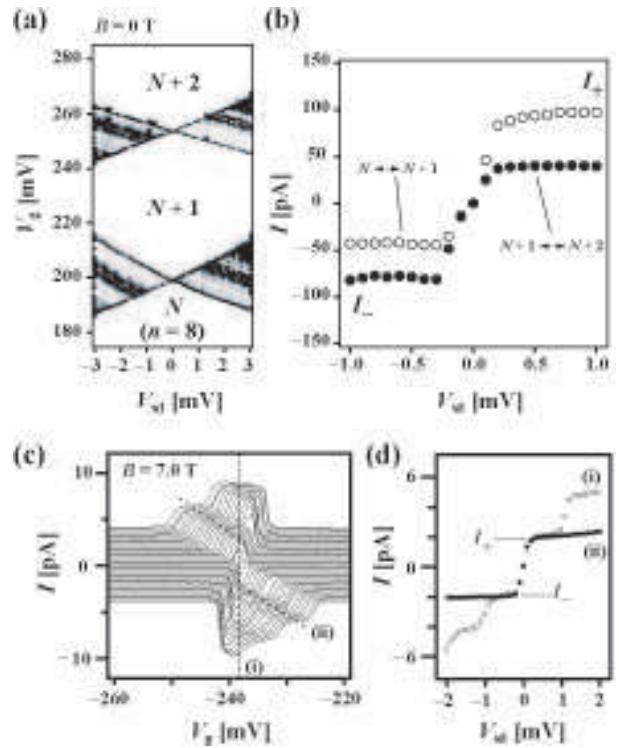}% 
\caption{\label{fig2} (a) Typical Coulomb diamonds ($n = 8$ to 10). (b) $I-V_{\text{sd}}$ characteristics of 
the intersection of adjacent Coulomb diamonds in (a); $N \leftrightarrow (N + 1)$ (open circles) and $(N + 1) \leftrightarrow (N + 2)$ (closed circles). (c) Typical Coulomb oscillations ($n = 0$ to 1) as a function of $V_{\text{g}}$ for $V_{\text{sd}}$ from $-2$ to 2 mV. A magnetic field of 7.0 T was applied perpendicular to the tube axis. Each peak is shifted for clarity. (d) $I-V_{\text{sd}}$ characteristics along the dashed lines (i) (open circles) and (ii) (closed circles) in (c).}
\end{figure}

We proceed to measure the saturation currents ($I_{+}$ and $I_{-}$) with and without magnetic fields for the region shown 
in Fig. 1(a). 
Figure 2(b) shows the typical $I-V_{\text{sd}}$ characteristics between $N$ $(n = 8)$- and $(N + 1)$- electron states (open circles) and between $(N + 1)$- and $(N + 2)$- electron states (closed circles) in Fig. 2(a) at $B = 0$ T. 
In this case, the ratio of saturation currents ($I_{+}/ I_{-}$) is obtained as $\sim$2 for $N \leftrightarrow (N +1)$ transition and $\sim$0.5 for $(N + 1) \leftrightarrow (N + 2)$ transition.
In magnetic fields, the Zeeman splitting of single-particle levels is expected, and new current steps appear for each individual single-particle level.  
The current steps of Coulomb staircases are very small in low magnetic fields and therefore the current ratio is not obtained by measuring the these staircases. However, the current steps can be clearly observed by measuring the Coulomb oscillations for large positive and negative biases. Figure 2(c) shows the typical Coulomb oscillations ($n = 0$ to 1) as a function of $V_{\text{g}}$ when $V_{\text{sd}}$ is changed from $-2$ to 2 mV at $B = 7.0$ T. Figure 2(d) (closed circles) shows the current values plotted along the first step of a Coulomb peak (dashed line (ii)) as a function of $V_{\text{sd}}$. The ratio of the saturation current can be easily determined even if the width of the current step is small. When the currents are plotted along the dashed line (i), which corresponds to the usual $I-V_{\text{sd}}$ characteristics at a fixed value of $V_{\text{g}}$, the obtained result is shown in Fig. 2(d) (open circles). $I_{+}/I_{-}$, which is obtained as $\sim$1 in this case, has the same values for each method. We have carefully investigated the spin states in the carbon nanotube quantum dot by performing these analyses in the magnetic fields from 0 to 10 T.

\begin{figure}
\includegraphics[width=8cm]{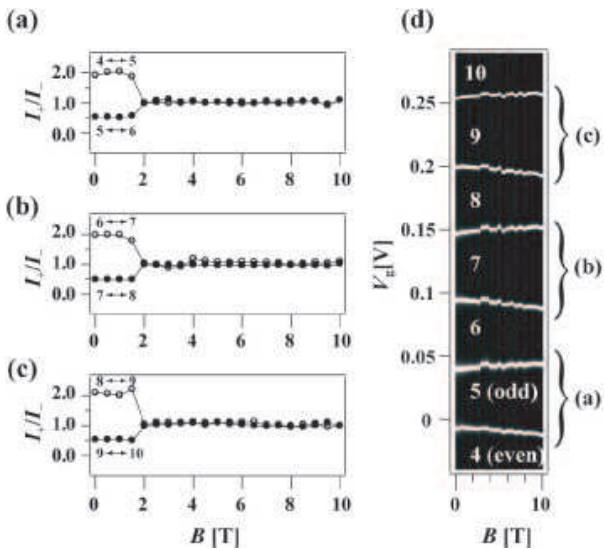}% 
\caption{\label{fig3} (a)$-$(c)Magnetic field dependence of the saturation current ratio ($I_{+}/I_{-}$) from $n = 4$ to 10 in Fig. 1(a) for the transition from even to odd-$N$ state (open circles) and for the transition from odd to even-$N$ state (closed circles).
(d) Magnetic field evolution of Coulomb peaks up to 10 T from $n = 4$ to 10 in Fig. 1(a). $V_{\text{sd}} = 0.1$ mV. 
The black-colored regions show the Coulomb blockaded regime, and the white-colored regions show the Coulomb peaks.}
\end{figure}

Figures 3(a)$-$(c) show the magnetic field dependence of $I_{+}/I_{-}$ from $n = 4$ to 10 in Fig. 1(a). 
Open circles indicate the current ratio for the transition from an even to an odd number of electrons and closed circles indicate that for the transition from an odd to an even number of electrons.
Up to $B \sim 1.5$ T, $I_{+}/I_{-}$ is $\sim$2 for all the even-to-odd transitions 
and $\sim$0.5 for all the odd-to-even transitions. 
These results indicate that the total spin changes from $S = 0$ to $S = 1/2$ when the number of electrons changes from even to odd. 
On the other hand, the total spin changes from $S = 1/2$ to $S = 0$ when the number of electrons changes from odd to even. 
Figure 3(d) shows the magnetic field evolution of Coulomb peaks up to 10 T from $n = 4$ to 10 in Fig. 1(a). The even-odd effect is again clearly demonstrated in evolution of the Coulomb peaks as a function of the magnetic field. All the peaks shift linearly as the magnetic field is increased, and the direction of the shift changes alternately as the number of electrons is increased by one unit~\cite{shell05}. This result is explained by the Zeeman splitting of each spin-degenerate single-particle level. The Zeeman splitting of the single-particle levels gives 
a $g$-factor of $\sim$2, and the total spin changes between 0 and 1/2 as the number of electrons is increased~\cite{APL01, Cobden}.
These results agree with the results of the current ratio in Fig. 3(a)$-$(c) at $B = 0$ T.
Then, at $B \sim 1.5$ T, the value of $I_{+}/I_{-}$ jumps to $\sim$1.0 and remains constant in the magnetic field larger than $B = 2.0$ T.
This effect is explained by the fact that the spin degeneracies of the single-particle levels are lifted by the Zeeman effect. In this situation, the argument that assumes the spin-degenerate levels is not applicable. Only one electron (spin-up or spin-down) can tunnel through the (spin-up or spin-down) Zeeman split level, and the tunneling rate is independent of the bias polarity. 
The Zeeman energy at $B = 2.0$ T is $\sim$230 $\mu$eV, which is significantly larger than the thermal energy in our experiment, where the current steps due to the Zeeman effect are clearly observed~\cite{explain01}.

\begin{figure}
\includegraphics[width=8cm]{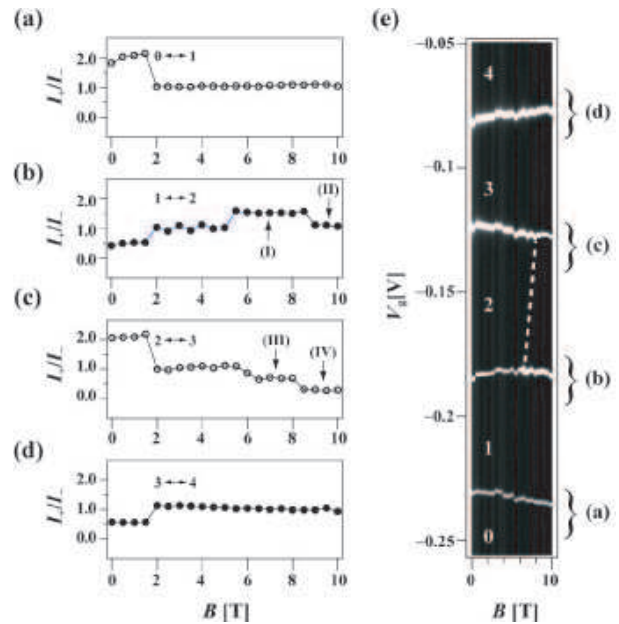}% 
\caption{\label{fig4} (a)$-$(d) Magnetic field dependence of $I_{+}/I_{-}$ from $n = 0$ to 4 in Fig. 1(a), 
for the transition from even to odd-$N$ state (open circles) and for the transition from odd to even-$N$ state (closed circles).
(d) Magnetic field evolution of Coulomb peaks up to 10 T from $n = 0$ to 4 in Fig. 1(a). $V_{\text{sd}} = 0.1$ mV. 
This plot is obtained by the same method that used to obtain the plot in Fig. 3(d). The dashed line connects the kink positions of two Coulomb peaks ($n = 1 \leftrightarrow 2$ 
and $2 \leftrightarrow 3$). This result is simillar to the four-electron shell-filling 
scheme; an internal spin-flip may occur in this line~\cite{shell02, Tans}.}
\end{figure}

Now, we consider the region of $n = 0$ to 4 shown in Fig. 1(a); this region shows a slightly more complicated behavior in magnetic fields.
Figures 4(a)$-$(d) show the magnetic field dependence of the saturation current ratio from $n = 0$ to 4 region. 
The $n = 0 \leftrightarrow 1$ and $3 \leftrightarrow 4$ transitions (Figs. 4(a) and (d), respectively) produce the same results as those in Fig. 3.

\begin{figure}
\includegraphics[width=8cm]{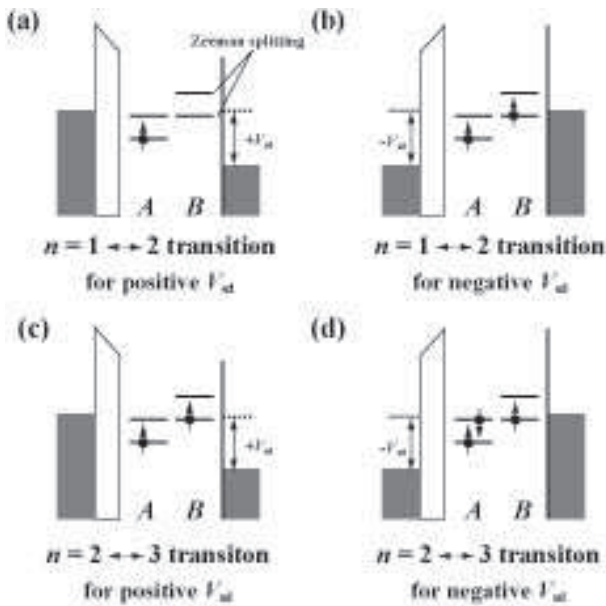}% 
\caption{\label{fig5}Mechanisims of the $n = 1 \leftrightarrow 2$ transition for (a) positive $V_{\text{sd}}$ and 
(b) negative $V_{\text{sd}}$ and the $n = 2 \leftrightarrow 3$ transion for (c) positive $V_{\text{sd}}$ and (d) negative $V_{\text{sd}}$ in magnetic fields in the single-particle model. $A$ and $B$ sites represent the ground state and the first excited state, respectively. Each single-particle levels have undergone Zeeman splitting. The right barrier is thinner in our experiment; therefore, tunneling current is limited only by the tunneling rate of the thicker left barrier (see text).}
\end{figure}

For the $n = 1 \leftrightarrow 2$ transition, $I_{+}/I_{-}$ changes from $\sim$1 to $\sim$1.5 at $B = 5.5$ T in Fig. 4(b), which theoretically corresponds to $3/2$ for $S_{N} = 1/2$ and $S_{N+1} = 1$. In magnetic fields, single-particle levels are lifted and no longer degenerate; consequently, the formula of the current ratio may not be applicable.
However, in this situation, two-electron spin states are degenerate due to the level crossings between the \textit{A}-spin-down state and the \textit{B}-spin-up state, as shown schematically in Figs. 5(a) and (b), and the formula of the current ratio may be applicable. Therefore, a higher-spin transition can be traced in region (I) in Fig. 4(b). 
When the degenerate levels between the \textit{A}-spin-down state and the \textit{B}-spin-up state differ slightly,  assuming that the crossing levels are lifted in Figs. 5(a) and (b), two electrons can enter the dot when $V_{\text{sd}}$ 
is positive, and two electrons can exit the dot when $V_{\text{sd}}$ is negative. Therefore, the value of the current ratio jumps to $\sim$1.0 again for larger magnetic fields (region (II)), and is consistent with the elementary value of 
the single-particle model. 
On the other hand, for $n = 2 \leftrightarrow 3$, $I_{+}/I_{-}$ changes from $\sim$1 to $\sim$0.67 at $B \sim 6.0$ T 
in region (III) in Fig. 4(c).
$I_{+}/I_{-}$ agrees with the theoretical value of $2/3$ for $S_{N} = 1$ and $S_{N+1} = 1/2$. 
At $B > 8.0$ T, $I_{+}/I_{-}$ jumps from $\sim$0.67 to $\sim$0.33 for $n = 2 \leftrightarrow 3$ in region (IV) in Fig. 4(c). In this case, three-electron spin states are no longer degenerate due to the Zeeman effect; therefore we can consider the simple single-particle model, assuming that the crossing levels are lifted in Figs. 5(c) and (d). Only one electron (spin-down) can enter the dot when the $V_{\text{sd}}$ is positive. On the other hand, three electrons can exit the dot when the $V_{\text{sd}}$ is negative; therefore, $I_{+}/I_{-}$ is equal to $1/3$ and is consistent with the elementary value of 
the single-particle model.
 
The above results indicate that at $B = 6$ T, the total spin changes from $0 \rightarrow 1/2 \rightarrow 1 \rightarrow 1/2 \rightarrow 0$ as $n$ increases.
Figure 4(d) shows the magnetic field evolution of the Coulomb peaks up to 10 T from $n = 0$ to 4 in Fig. 1(a). This behavior is similar to the four-electron shell-filling scheme, as reported previously~\cite{shell02}.
These results are consistent with the results of the current ratio shown in Fig. 4(a)$-$(d). We show that the higher-spin transitions ($S = 1$) in magnetic fields can be traced using this method.

In conclusion, we have determined the total spin and its magnetic field dependence by using the ratio of the saturation currents for positive and negative biases. Higher-spin transitions in magnetic fields can be traced by this method when the single-particle levels are degenerate; then, the single-particle model cannot be applied in the sigle-electron transport such that shown in Fig. 5. Furthermore, these results indicate that the spin-up and spin-down states of a single-particle level that has undergone Zeeman splitting have an equivalent tunneling rate in the SWCNT quantum dot. The experimental results are consistent with those of the total-spin study due to the magnetic field evolution of Coulomb peaks, as reported earlier. This method can be used to determine the total spin in quantum dots.  

We highly acknowledge the discussions and suggestions of Y. Utsumi of RIKEN. We also thank A. Furusaki of RIKEN and H. Akera of Hokkaido University for useful discussions and comments. This study was supported in part by the Special Postdoctoral Researchers Program of RIKEN and by the Grant-in-Aid for Young Scientists (B) (18710124) from the Ministry of Education, Culture, Sports, Science and Technology, Japan.


\begin{thebibliography}{30}
\bibitem{explain02}
\bibinfo{note}{In this case, the current ratio is given by $(\varGamma_{\text{L}} + 2 \varGamma_{\text{R}})/(2 \varGamma_{\text{L}} + \varGamma_{\text{R}})$~\cite{Cobden}. For $\varGamma_{\text{L}} \ll \varGamma_{\text{R}}$, $I_{+}/I_{-} \to 2$.}
\bibitem{explain01}
\bibinfo{note}{In fact, we can observe the influence of the Zeeman effect in the $I-V_{\text{sd}}$ or $I-V_{\text{g}}$ 
characteristics at a lower magnetic field ($B \sim 1.0$ T). However, the widths of the new current steps due to the Zeeman effect are very small; therefore, we define clear Zeeman splittings (current steps) as nearly twenty times larger than the thermal energy.}
\end{thebibliography}
\end{document}